# Modernizing use of regression models in physics education research: A review of hierarchical linear modeling

Ben Van Dusen and Jayson Nissen

*Department of Science Education, California State University Chico, Chico, California, 95929, USA*

[This paper is part of the Focused Collection on Quantitative Methods in PER: A Critical Examination.] Physics education researchers (PER) often analyze student data with single-level regression models (e.g., linear and logistic regression). However, education datasets can have hierarchical structures, such as students nested within courses, that single-level models fail to account for. The improper use of single-level models to analyze hierarchical datasets can lead to biased findings. Hierarchical models (also known as multilevel models) account for this hierarchical nested structure in the data. In this publication, we outline the theoretical differences between how single-level and multilevel models handle hierarchical datasets. We then present analysis of a dataset from 112 introductory physics courses using both multiple linear regression and hierarchical linear modeling to illustrate the potential impact of using an inappropriate analytical method on PER findings and implications. Research can leverage multi-institutional datasets to improve the field's understanding of how to support student success in physics. There is no *post hoc* fix, however, if researchers use inappropriate single-level models to analyze multilevel datasets. To continue developing reliable and generalizable knowledge, PER should use hierarchical models when analyzing hierarchical datasets. The Supplemental Material includes a sample dataset, R code to model the building and analysis presented in the paper, and an HTML output from the R code.

## I. INTRODUCTION

Early work in physics education research (PER) focused on student conceptual change and it remains a popular area of study [1]. To measure conceptual change, physics education researchers often analyze students' pretest and post-test scores on research-based assessments (RBAs) [2], such as the Force Concept Inventory (FCI) [3]. To identify the impact of a course transformation, physics education researchers commonly compare student data collected across course contexts (e.g., traditional and transformed courses). Including data from multiple contexts also has the benefit of improving a study's statistical power and the generalizability of its findings. Many of the most cited publications in the PER literature rely on data collected from multiple courses and institutions [4–7]. Data from multiple contexts introduces a hierarchical structure into the data where student data (level 1) nests within course data (level 2). This nesting can include additional levels, such as departments (level 3) and institutions (level 4). When included in a multilevel regression model, the structure of a hierarchical dataset can provide additional information. The assumption of independence, which is central to many statistical analyses [e.g., multiple linear regression (MLR) and analysis of variance (ANOVA)], is violated by connections between data points within a hierarchical dataset. The violation of the assumption of independence can bias the results from statistical models [8,9]. The assumption of independence requires that there are no student groupings that are not accounted for in a model. Hierarchical models account for the nested nature of the data and do not require the assumption of independence to be met for the results to be reliable.

Many disciplines generate hierarchical datasets. While researchers in each discipline use similar hierarchical models to analyze their data, they refer to the models by different names. For example, hierarchical models are referred to as multilevel linear models in sociology, mixed-effects models and random-effects models in biometrics, random-coefficient regression models in econometrics, and covariance components models in the statistical literature [10]. In this publication, we will use hierarchical linear models (HLM) because it is the nomenclature education researchers commonly use for hierarchical models.

The purpose of this article is to assist researchers in identifying and applying the regression analysis techniques best suited to their data and research questions. We will accomplish this purpose in three sections: (i) Motivation— we review PER's historical use of regression analyses,

(ii) theory—we discuss the theoretical advantages and disadvantages of three common techniques for dealing with hierarchical data, and (iii) application—we examine the practical implications of using MLR vs HLM.

## II. MOTIVATION

Physics education researchers commonly use regression models to investigate a wide range of phenomena including conceptual learning, attitude development, drop, fail, and withdrawal rates, and likelihoods of passing a class. These models isolate the impact of a variable of interest while controlling for other variables. For example, to investigate the impact of research-based teaching practices on women, a researcher might create a linear regression model of post-test scores that controls for student pretest scores and gender. To determine the different regression models commonly used in PER, we reviewed publications in Physical Review Physics Education Research. We performed our literature search on 3/27/18 using the "All Field" search tool on the Physical Review PER website [11]. Our search of the PER literature included both the Physical Review PER and Physical Review Special Topics PER archives. We did not include PER Conference proceedings or the American Journal of Physics since their search functions did not support our query. We searched for the terms "linear regression," "hierarchical linear model," "multi-level model," and "multilevel model." We performed an identical search of the broader education literature using the Sage Journal's "anywhere" search tool of the journals they classify as being "education." Examples of Sage's education journals include Educational Researcher and American Education Research Journal.

A single-level linear regression called multiple linear regression (MLR) was the most commonly used type of regression model. 43 publications in Physical Review PER and 432 publications in Sage education journals mentioned linear regression. It was not surprising that there were more publications that mentioned linear regression in Sage education journals given that they included multiple journals. Our search for hierarchical linear model and its synonyms in Physical Review PER returned only 2 publications and in Sage education journals returned 196 publications. While Sage journals had $\sim$10$\times$ as many publications that mentioned linear regression they had $\sim$100$\times$ as many publications that mentioned hierarchical linear model and its synonyms. Of the two publications in Physical Review PER that mentioned hierarchical linear model, the first mentioned HLM as a possible method of analysis but did not use it [12]. The second publication stated that they performed both HLM and linear regression, but because they had similar results they only included the linear regression results in the publication [13]. Independent of our literature search, we identified two Physical Review PER articles [14,15] that used HLM but did not show up in our search because they referred to HLM using the less common nomenclature of hierarchical-model analysis and hierarchical linear regression. The more common use of HLM in Sage education journals likely results from education researchers starting to use HLM thirty years ago [16] and its increased use as access to computing power and larger datasets improved.

Despite not using HLM, physics education researchers have often analyzed hierarchical datasets [4,6,17–19]. We propose that the PER community seldom used HLM due to a lack of knowledge about both HLM and the potential for traditional methods to bias findings. While resources existed for applying HLM in different disciplines [8,20,21], none existed in PER. In Secs. III and IV, we address this limitation by outlining the theoretical differences between MLR and HLM and then applying both methods to a hierarchical dataset from 112 physics courses to illustrate the impact that the selection of an inappropriate modeling technique can have on findings.

## III. THEORY

### A. Sampling

Most sampling designs assume independence between each measurement. That is, each student is independent from the others, and all students have an equal chance of being selected. Data collected from a single course is more likely to meet the assumption of independence since it limits the contexts and opportunities for the data to be hierarchical. Data can still be hierarchical within a single course, however, if the data nests within individuals or groups of students. For example, in a course with three unit exams and a final, the four exam scores are nested within each student and create a hierarchical dataset. When analyzing hierarchical datasets, however, researchers often fail to consider if their data meet the assumption of independence and simple proceed as if it does. While erroneously making this assumption will not always lead to biased results, it calls into question the validity of the claims they can support [8,9,22].

Even if researchers collect an abundance of pertinent data about the courses they study, those courses include features that cannot be explicitly accounted for. For example, a researcher may know which introductory physics courses in their dataset were algebra and calculus based, if it was the fall or spring semester, who the instructors were, how much experience the instructors had teaching the course, and what the student to instructor ratio was, but there will be other factors that are harder to measure, such as whether a class was held at 8 a.m. leading some students to show up late, a particularly nasty flu on campus that semester, or whether the instructor was preoccupied preparing their tenure packet. These known and unknown course features interact to either improve or depress student performance within a course. The impact of the unique features of courses on students leads to data clustering within courses and violates the assumption of independence [22]. Given that numerous student-, course-, and institution-level features can influence student performance, researchers should use a method that accounts for the nesting of student data when analyzing hierarchical datasets. Table I shows example variables that a researcher may want to account for within each hierarchical level.

TABLE I. Factors at each hierarchical level that may affect students' concept inventory post-test score.

| Hierarchical level | Example of hierarchical level | Example variables |
|---|---|---|
| Level 3 | Institution level | Public vs private 2-year vs 4-year Teaching intensive Curriculum Small-group work |
| Level 2 | Course level | Learning assistants Instructor experience Class mean pretest Gender Race or ethnicity Major |
| Level 1 | Student level | Math preparation SAT scores Pretest score Self-efficacy |

### B. Modeling

Social scientists have used three primary methods to analyze hierarchical datasets: disaggregation, aggregation, and HLM. To examine the implications of using these three methods, we will apply each of them to a sample multilevel dataset (shown in Table II) to predict student post-test performance on the FCI. The dataset includes information

TABLE II. Sample multilevel dataset.

| Student (Level 1) | Course (Level 2) | Student SAT verbal (Level 1) | FCI post-test (Level 1) |
|---|---|---|---|
| 1 | 1 | 500 | 50 |
| 2 | 1 | 600 | 70 |
| 3 | 2 | 550 | 40 |
| 4 | 2 | 650 | 60 |
| 5 | 3 | 600 | 30 |
| 6 | 3 | 700 | 50 |
| 7 | 4 | 650 | 20 |
| 8 | 4 | 750 | 40 |
| 9 | 5 | 700 | 10 |
| 10 | 5 | 800 | 30 |

about the students (level 1) and the courses they are nested in (level 2).

#### 1. Disaggregation

In reviewing the literature, we found that physics education researchers commonly used disaggregation to analyze hierarchical datasets. Disaggregation ignores all nesting in the dataset and either excludes level-2 data or treats it as level-1 data. For example, many investigations of gender differences in introductory physics courses disaggregated data across multiple courses and largely ignored course-level data to obtain large enough samples to support their investigations (e.g., Refs. [19,23–25]). In another example, Nissen and Shemwell [26] investigated gender differences in student's experiences. Their data nested multiple experiences (level 1) within students (level 2). They disaggregated the data by treating gender as a level-1 variable instead of using HLM to account for the experiences nesting within the students. Linear regressions of disaggregated data fail to account for course-level variance in hierarchical datasets and violate the method's assumptions of independence [20,27,28]. This can lead to bias in findings, an underestimation of the standard error, and artificially small $p$ values [8,22].

A single-level linear regression of the disaggregated data from our example data found that student SAT verbal scores had a small negative association ($-0.067$) with FCI post-test scores [Fig. 1(b)]. In other words, the model predicted students with higher SAT verbal scores would have slightly smaller FCI post-test scores. This slight negative association came from the small negative trend between SAT verbal scores and FCI post-test scores.

#### 2. Aggregation

Aggregation collapses level-1 data into level-2 data. In the case of our example dataset, we averaged the student-level variables of SAT verbal scores and FCI post-test scores within each course to create class mean SAT verbal scores and class mean FCI post-test scores, which is shown in Fig. 1(c). A single-level linear regression of the aggregated course-level data finds that course mean SAT verbal scores have a negative association ($-0.2$) with course mean FCI post-test scores [Fig. 1(c)]. In this case the aggregation model identified a similar negative association between course mean SAT verbal score and FCI post-test scores as the disaggregation model, but the negative coefficient was 3 times larger.

In reviewing the literature, we found that physics education researchers sometimes use aggregation to analyze hierarchical datasets. Hake [4] aggregated students across courses to investigate differences in student conceptual learning in active engagement and lecture-based courses.

By aggregating student-level data into course-level data, researchers lose all information about student-level variability. Aggregation treats each group equally and fails to account for precision of the measurements. For example, a class with 10 students can have an equally large impact on a model as a class of 500 students. By eliminating all lower-level data, aggregation meets the assumption of independence. As such, aggregation can be an appropriate method if one is only interested in course-level processes. Aggregation can cause several errors, however, when investigating the impact of student-level factors. One potential error is a shift in meaning. For example, our aggregation model that examines the course mean SAT score can say something about the impact of a course context, but does not address the impact of an individual student's SAT verbal score. The majority of the variability in student scores comes from the student level, not the course level [8,10]. Examining only the impact of course-level variables on student performance ignores most of the variability in outcomes and can distort relationships between variables.

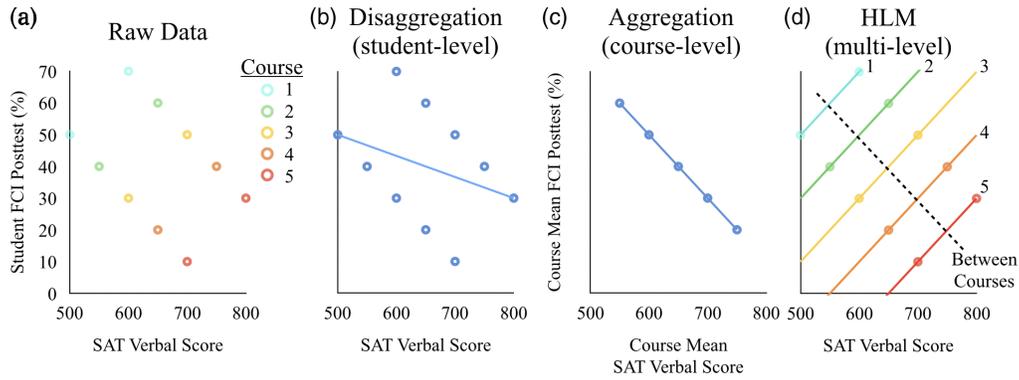

FIG. 1.   Example dataset of the relationships between SAT verbal and Force Concept Inventory (FCI) post-test scores analyzed using 4 different methods. (a) A plot of the raw data without any model. (b) Disaggregation, which ignores any course groupings and analyzes all of the student-level data. (c) Aggregation, which averages student-level data to create course-level data and then analyzes them. (d) Hierarchical linear modeling (HLM), which analyzes student- and course-level data simultaneously. We adapted the figure from an example in Snijders and Bosker [22] and Woltman et al. [8].

Researchers will sometimes combine aspects of the disaggregation and aggregation techniques into a single MLR model. For example, in Kost and colleague's examination of the gender gap in introductory physics courses [23], they examined data from 2099 students in seven first-semester mechanics physics courses. They disaggregated their data and analyzed it using multiple linear regression models. To account for the fact that their dataset included data from multiple semesters, they included a dummy variable for each semester, which allowed the model to begin to account for differences in average scores across semesters. While including a variable for each group in a dataset can improve the quality of a single-level model examining hierarchical data, it has significant limitations. For example, to include level-2 dummy variables in the MLR developed in Sec. IV would require the inclusion of 111 additional variables (1 less than the number of courses in the dataset) and would limit our findings about the impact of course features on the model.

### 3. Hierarchical linear modeling

Hierarchical linear modeling (HLM) does not have the assumption of independence (unlike disaggregation and aggregation) and was developed specifically to analyze hierarchical datasets [8,10]. HLM accomplishes this by creating unique regression models for each course to model student-level variables and examines difference between sections to model course-level variables. Analyzing the data from Table II with HLM shows a positive association (0.2) between student SAT verbal and FCI post-test scores and a negative association ($-0.2$) between class mean SAT Verbal and FCI post-test scores [Fig. 1(d)]. In other words, the model predicts the highest FCI post-test scores for students with high SAT verbal scores in classes with low average SAT verbal scores.

HLM provides the advantages of both disaggregation and aggregation without introducing their shortcomings [8,10] making it the preferred method for handling multi-level data. To accomplish this, HLM creates a set of equations that nests level-2 equations within level-1 equations. Equations (1)–(3) illustrate a basic HLM model with one predictor variable at level 1 ($Lvl1variable$) and one at level 2 ($Lvl2variable$). In HLM equations the $i$ subscript denotes level-1 units (e.g., students) and the $j$ subscript denotes level-2 units (e.g., courses).

**Level-1 Equation:**

$$Outcome_{ij} = \beta_{0j} + \beta_{1j}Lvl1variable_{ij} + r_{ij}. \quad (1)$$

**Level-2 Equations:**

$$\beta_{0j} = \gamma_{00} + \gamma_{01}Lvl2variable_j + u_{0j}, \quad (2)$$

$$\beta_{1j} = \gamma_{10} + u_{1j}. \quad (3)$$

HLM creates a level-2 equation for each level-1 coefficient, which equations (2) and (3) illustrate. In contrast,

the coefficients in MLR models are all values. In HLM, a researcher can allow for variation across groups for each level-1 variable by including a new variance term ($u$) in the level-2 equations; this is denoted by $u_{0j}$ and $u_{1j}$ in equations (2) and (3). This variance term allows the relationship with the associated level-1 variable to differ across groups. For example, a researcher may want to allow the impact of student gender on test scores to vary across courses because they believe that it will have a larger impact in some course contexts than others. Level-1 variables are called random slope variables if they have a variance term (e.g., $u_{1j}$) associated with them and are called fixed slope if they do not. The coefficients in the level-2 models (represented by $\gamma$) are referred to as fixed effects because they are not allowed to vary across groups while the $u$ terms are called random effects because they account for the random differences between the groups. While including additional variance terms allows models to be more flexible, they also require more statistical power and are only recommended if they improve the fit of a model.

The nesting of level-2 equations within the level-1 equation allows the model to account for variance at both levels simultaneously. By accounting for variance at multiple levels, HLM can test for dependencies among level-1 units (students) within each level-2 unit (course). These coefficients are generated from a combination of information from each individual course's data and overall information across all of the courses. To combine information from the individual courses and the grand mean, HLM uses Bayes "shrinkage estimators" that take into account several factors about each course. Of the factors, the sample size within the course is the most important [29].

By developing a model with no predictor variables, called the unconditional model, researchers can use HLM to calculate the variance at both the student level [variance $(r_{ij}) = \sigma^2$ = within group variance] and the class level [variance $(u_{0j}) = \tau^2$ = between group variance]. The percentage of the total variance ($\tau^2 + \sigma^2$) attributed to the between group variance ($\tau^2$) is known as the *intraclass correlation coefficient* (ICC). The ICC quantifies the share of the differences in student scores caused by differences in students versus differences in courses. If the data do not significantly vary at the class level (ICC < 5%), then a single-level and hierarchical model will likely provide similar findings and MLR may be appropriate. However, the only way to know that HLM and MLR provide similar results is to run both analyses and compare them.

$$\text{ICC} = (\text{between group variance})/(\text{total variance}) \quad (4)$$

$$= \text{var.}(u_{0j})/[\text{var.}(u_{0j}) + \text{var.}(r_{ij})] \quad (5)$$

$$= \tau^2/(\tau^2 + \sigma^2). \quad (6)$$

## C. HLM assumptions

While HLM has fewer assumptions than standard regression analysis, failure to meet model assumptions can still lead to misrepresentations of the relations in the data and $p$ values [22]. HLM's four assumptions are (i) *linearity*—the dependent variables should vary linearly with the explanatory variable, (ii) *independence of residuals*—level-1 and level-2 residuals are uncorrelated, (iii) *homoscedasticity*—residuals should be distributed in a normal and homoscedastic (i.e., similarly across the range of outcome values) manner, and (iv) *random intercepts*—variables must be allowed to have random intercepts [22]. There are a variety of methods by which the assumptions of HLM can be checked, many of which involve visual inspection of residual plots [30]. We will demonstrate some of these methods for checking HLM assumptions in Sec. IV where we analyze an HLM model. There are methods for fixing HLM models that violate assumptions. For example, if the assumption of linearity is violated, performing a transformation of a variable (e.g., squaring or log transformation) may fix it. For more information about the assumptions of HLM and how to check them, we recommend referencing Snidjers and Bosker's book on multilevel analysis [22] or Loy's dissertation on diagnostics for mixed or hierarchical linear models [31].

Because HLM allows for variance at all levels, it creates a separate regression model for each course that accounts for its unique features. It then combines the course models to create a set of coefficients that describe the larger dataset. This hierarchical structure provides HLM several advantages over MLR. HLM can accommodate missing data (at level 1), small and/or discrepant group sample sizes [8], and nonsphericity (i.e., a sample is from a population in which either variances are not equal or correlations are nonzero) [32].

The primary disadvantage of HLM for researchers is that it requires more statistical power (i.e., larger sample sizes) than MLR. HLM also requires sufficient samples at each level of the model. Heuristics for the minimum sample at level 2 vary by citation and can be as large as 50, but Maas and Hox [33] found that level-2 sample sizes as small as 10 can produce unbiased estimates.

## IV. APPLICATION

In Sec. III, we used example data to highlight the theoretical differences between MLR and HLM. While this shows how HLM can create more accurate models when analyzing hierarchical datasets, it fails to demonstrate whether using MLR and HLM lead to meaningfully different conclusions when analyzing real-world physics student data. To test how using these two analytical techniques may lead to different conclusions about physics student learning, we reanalyzed a dataset from a prior investigation [34] using both MLR and HLM. In addition to illustrating how analysis methods can influence findings, this section serves to demonstrate the appropriate steps in developing HLM models.

The original investigation disentangled the relationships between learning assistants (LAs), the collaborative learning activities that they support, and student learning. Learning assistants are undergraduate students who, through the guidance of weekly preparation sessions and a pedagogy course, facilitate discussions among groups of students in a variety of classroom settings that encourage active engagement [35]. To isolate the relationship between LAs and student learning, Herrera *et al.* [34] compared student learning across courses in which instructors reported using lecture, collaborative learning *without* LAs, or collaborative learning *with* LAs.

### A. Research question

In our reanalysis of the data using MLR and HLM, we interpret our models to answer the following question:
- How does the use of MLR or HLM impact findings about the efficacy of physics course pedagogical practices?

### B. Methods

#### 1. Data collection and preparation

Our study analyzes course and student Force Concept Inventory (FCI) [3] and the Force and Motion Conceptual Evaluation (FMCE) data from the Learning About STEM Student Outcomes (LASSO) platform. The LASSO platform is hosted on the LA Alliance website [36] and collects large-scale, multi-institution data by hosting, administering, scoring, and analyzing pretest and post-test research-based assessments online.

To clean the data, we removed assessment scores from students who took less than 5 min on the assessment or completed fewer than 80% of the questions. Five minutes provided a reasonable minimum amount of time for a student to complete the CI while reading and answering each question. We removed courses with less than 40% student participation on either the pretest or post-test from the data. Lack of student participation led to the removal of 44 courses from the dataset. Of the 44 courses removed, 8 had no pretests, 11 had no post-tests, and 25 had less than 40% participation on the post-test. The filters for time and completion removed 269 pretest scores and 398 post-test scores. This led to the removal of 258 students from the dataset who did not have either a pretest or a post-test score. Table III shows the number of students, courses, and students remaining in the dataset after each step of data cleaning. The final dataset included data from 5959 students in 112 courses at 17 institutions with missing data for 15% of the pretests and 30% of the post-tests. This resulted in 55% of the responses having matched pretest and post-test which falls in the middle of the 30% to 80% range of matched data reported in the PER literature [37]. We calculated pretest and post-test scores using the total percentage correct of all the items on the assessment.

TABLE III. Size of the dataset after each step of filtering.

|              | Initial | Course participation | Time and completion |
|--------------|---------|----------------------|---------------------|
| Institutions | 20      | 17                   | 17                  |
| Courses      | 156     | 112                  | 112                 |
| Students     | 8329    | 6217                 | 5959                |

After cleaning the data, we used hierarchical multiple imputation (HMI) with the hmi [38] and mice [39] packages in R-Studio V. 1.1.456 [40] to address missing data. HMI maximizes statistical power by addressing missing data while taking into account the hierarchical structure of the data [41–44]. HMI addresses missing data by (i) imputing each missing data point $m$ times to create $m$ complete datasets, (ii) independently analyzing each dataset, and (iii) combining the $m$ results using standardized methods [45]. Multiple imputation does not produce specific values for missing data; rather, it uses all the available data to produce valid statistical inferences [43].

PER seldom uses multiple imputation [46–48] and prefers to use complete-case analysis [37], where researchers discard cases that do not include both a pretest and a post-test. However, research indicates that multiple imputation leads to better analytics than traditional methods such as complete-case analysis [41]. While traditional methods for addressing missing data are computationally simple, they have likely biased findings and measures of statistical significance [37,42]. A significant driver of bias in student gains using complete case analysis comes from differential participation rates across physics performance groups. Students who perform well in a physics course tend to have high pretest scores, earn higher grades in the course, and participate at higher rates on low-stakes assessments [48]. The bias in participation leads samples to underrepresent low-performing students. Multiple imputation can use data that is linked to student performance (i.e., pretest and post-test scores) to mitigate some of the bias in sampling when imputing values for students with partial data. While no research has empirically established how much of the bias in gains is accounted for by pretest or post-test scores, other performance indicators, such as final grades in a course, have been shown to account for bias in gains [37]. Extensive research indicates that in almost all cases MI is both more accurate and more statistically powerful than complete-case analysis [43,49].

Our HMI model included the following variables: concept inventory used, student pretest and post-test scores and durations, whether it was a student's first time taking the course, student race or ethnicity, student gender, and the type of instruction in the course. We expected these variables accounted for some of the bias in missing data, and that MI produced more accurate and reliable results than complete-case analysis. The data collection platform (LASSO) provided complete datasets for the concept inventory variables, student demographics, and instruction type. The 45% rate of missing data (15% on pretests plus 30% on post-tests) in this study was within the normal range for PER studies [48]. The HMI produced 10 imputed complete datasets. We analyzed all 10 imputed data sets and combined the results. Final results were created by averaging the test statistics (e.g., model coefficients) and using Rubin's rules to combine the standard errors for these test statistics [44]. Rubin's rules combines the standard errors using both the within-imputation variance and the between-imputation variance with a weighting factor for the number of imputations used. All MI assumptions were satisfactorily met for all of the MI analyses. For readers seeking more information on MI, Schafer [44] and Manly and Wells [43] present overviews of MI and we provide a companion article on missing data and multiple imputation in pretest and post-test data in PER in this special issue [37].

### 2. Model development

We used the mitml [50] and lme4 [51] packages in R to analyze the 10 imputed datasets and pool the results. Figure 2 shows our work flow for the data collection and analysis. The MLR and HLM models we developed included identical variables, with the exception of a variable ($u_{0j}$) that allowed the HLM model to account for course-level groupings of student data. The Supplemental Material [52] includes data and R code for running the MLR and HLM models on one of the 10 imputed datasets.

To investigate student learning, we developed a set of HLM models to predict student gains (*post-test−pretest*). The use of gains as the outcome variables is functionally equivalent to using the post-test score as the outcome variable in our models, but leads to findings that are more easily interpretable. As the mean scores for the pretest (36%) and post-test (55%) in our dataset indicate that the floor and ceiling effects were limited, no transformation of the data was required prior to analysis [53]. In our analysis, pretest and gains are represented as a percentage correct. Our HLM models were all two-level with student data in the first level and course data in the second level. We chose not to develop three-level models with institutional data in the third level because we were limited by the number of institutions in our dataset ($n = 17$) and because our exploratory analysis did not indicate that the inclusion of a third level changed the findings.

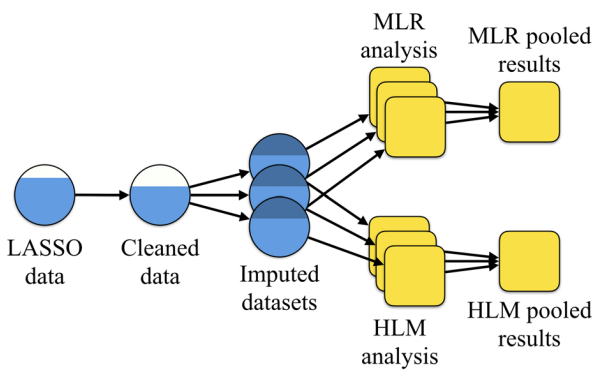

FIG. 2. Work flow for collecting, preparing, and analyzing the data using MLR and HLM. In the circles, blue represents data, white represents missing data, and navy blue represents imputed data. Squares represent results.

We developed our models, shown in Table IV, for student gain scores through a series of additions of variables. Model 1, the unconditional model, predicted student gains without level-1 or level-2 predictor variables. The unconditional model allowed us to calculate the intraclass correlation coefficient (ICC) by comparing the course- and student-level variance in the HLM version of model 1. The ICC indicated course-level effects accounted for 13% of the variation in student gains and fell above the heuristic threshold of 5%, indicating the need for analysis with HLM.

We developed models 2–6 by the incremental addition of predictor variables that either improved the model fit or served to answer the research questions. With the exception of model 2, each model introduced 1 new variable. Model 2 introduced the variables for classes that use LAs ($LA_j$) and collaborative learning without LAs ($CollNoLA_j$) at the same time as they were the two variables of interest in our study. We evaluated each model on whether the inclusion of its additional variable improved the model's goodness of fit. Goodness of fit can be assessed through the examination of several different statistics such as Akaike information criterion (AIC), Bayesian information criterion (BIC), and variance explained. As there is currently no agreed upon way to pool the AIC or BIC statistics for multilevel models across multiply imputed datasets, we used variance explained to select our final model. Because we want the simplest model that can account for the most variance we only included variables that could explain at least 1% of the student- or course-level variance. Table IV shows the equations used in each model and the variances at each level. Additional variables examined included student pretest score ($StudentPre_{ij}$), course mean pretest score ($CoursePre_j$), a random effect variable ($u_{1j}$), and an instrument variable ($FMCE_j$). $StudentPre_{ij}$ was group mean centered and $CoursePre_j$ was grand mean centered. Details about centering are included in the subsequent Sec. IV B 3.

We concluded that model 3 was our best model. While the inclusion of $StudentPre_j$ in model 3 decreased the level-2 explained variance by 3%, it improved the level-1 explained variance by 16.2%. The HLM version of model 3 is specified in Eqs (7)–(9) and the MLR version is specified in Eq. (10). Each of the variables included in models 4, 5, and 6 failed to improve the fit of the model by at least 1% at either the student or course level.

TABLE IV. Model development with final estimation of variance components (Var.) and the percent variance explained (% expl.) compared to model 1, the unconditional model. The inclusion of additional variables (shown in bold) in models 4, 5, and 6 failed to improve the percent explained variance by more than 1% at either level. We used model 3 to compare MLR and HLM.

| Model | Level | Equation | Var. | % expl. |
|-------|-------|----------|------|---------|
| 1 | 1 | $gain_{ij} = \beta_{0j} + r_{ij}$ | 409.8 | $\cdots$ |
|   | 2 | $\beta_{0j} = \gamma_{00} + u_{0j}$ | 60.6 | $\cdots$ |
| 2 | 1 | $gain_{ij} = \beta_{0j} + r_{ij}$ | 409.8 | 0% |
|   | 2 | $\beta_{0j} = \gamma_{00} + \gamma_{01} * \boldsymbol{LA_j} + \gamma_{02} * \boldsymbol{CollNoLA_j} + u_{0j}$ | 56.6 | 6.6% |
| 3 | 1 | $gain_{ij} = \beta_{0j} + \beta_{1j} * \boldsymbol{StudentPre_{ij}} + r_{ij}$ | 343.5 | 16.2% |
|   | 2 | $\beta_{0j} = \gamma_{00} + \gamma_{01} * LA_j + \gamma_{02} * CollNoLA_j + u_{0j}$ $\beta_{1j} = \gamma_{10}$ | 58.8 | 3.0% |
| 4 | 1 | $gain_{ij} = \beta_{0j} + \beta_{1j} * StudentPre_{ij} + r_{ij}$ | 340.3 | 17.0% |
|   | 2 | $\beta_{0j} = \gamma_{00} + \gamma_{01} * LA_j + \gamma_{02} * CollNoLA_j + u_{0j}$ $\beta_{1j} = \gamma_{10} + \boldsymbol{u_{1j}}$ | 58.8 | 3.0% |
| 5 | 1 | $gain_{ij} = \beta_{0j} + \beta_{1j} * StudentPre_{ij} + r_{ij}$ | 343.5 | 16.2% |
|   | 2 | $\beta_{0j} = \gamma_{00} + \gamma_{01} * LA_j + \gamma_{02} * CollNoLA_j + \gamma_{03} * \boldsymbol{CoursePre_j} + u_{0j}$ $\beta_{1j} = \gamma_{10}$ | 59.4 | 2.0% |
| 6 | 1 | $gain_{ij} = \beta_{0j} + \beta_{1j} * StudentPre_{ij} + r_{ij}$ | 343.5 | 16.2% |
|   | 2 | $\beta_{0j} = \gamma_{00} + \gamma_{01} * LA_j + \gamma_{02} * CollNoLA_j + \gamma_{03} * \boldsymbol{FMCE_j} + u_{0j}$ $\beta_{1j} = \gamma_{10}$ | 59.4 | 2.0% |

**HLM Level-1 Equation (Model 3):**

$$Gain_{ij} = \beta_{0j} + \beta_{1j} StudentPre_{ij} + r_{ij}. \qquad (7)$$

**HLM Level-2 Equations (Model 3):**

$$\beta_{0j} = \gamma_{00} + \gamma_{01} LA_j + \gamma_{02} CollNoLA_j + u_{0j}, \qquad (8)$$

$$\beta_{1j} = \gamma_{10}. \qquad (9)$$

**MLR Equation (Model 3):**

$$Gain_i = \beta_0 + \beta_1 StudentPre_i + \beta_2 LA_i$$
$$+ \beta_3 CollNoLA_i + \epsilon_i. \qquad (10)$$

### 3. Centering

Researchers can choose to leave variables uncentered or center them using either group mean centering or grand mean centering. Group mean centering centers a student variable ($i$) within a related course variable ($j$). In other words, group mean centering transforms the student variable to be a measure of how much that student variable differed from the course's mean. In contrast, grand mean centering centers the individual student variable ($i$) about the average for all students on that variable. Centering variables can make the model easier to interpret and can change the value and meaning of the coefficients in the model. The type of centering researchers use depends on their research questions. A detailed discussion of all of the reasons, costs, and affordances for centering variables exceeds the scope of this article, but we provide a brief justification of our choices and recommend interested readers review the literature on centering in HLM [54–57].

For ease of interpretation, we group mean centered $StudentPre_{ij}$. This has two effects on the models. First, the intercept $\beta_{0j}$ represents the predicted gain for a student who had the average pretest score in their course (i.e., group). Second, the coefficient for $StudentPre_{ij}$ informs the relationship between an above (or below) average pretest in a course and the predicted gain. We used group mean centering because that is generally recommended in the literature. This is particularly useful in our case where an uncentered model predicts values for a pretest of zero, which is unlikely [57] and more difficult to interpret. Group mean centering $StudentPre_{ij}$ kept the intercept consistent across the models and produced models with lower variance.

Group mean centering $StudentPre_{ij}$ can hide differences between courses because it centers all of the courses around their own mean pretests. Therefore, we included the level-2 variable $CoursePre_{ij}$ in model 5 to inform the extent to which predicted gains differed across courses with different pretest distributions. We grand mean centered $CoursePre_{ij}$ to maintain the models as predicting the gain for a student in a course with an average mean pretest score since a course with a mean pretest score of zero is highly improbable. The R code included in the Supplemental Material [52] shows the specifics of how the centering was performed.

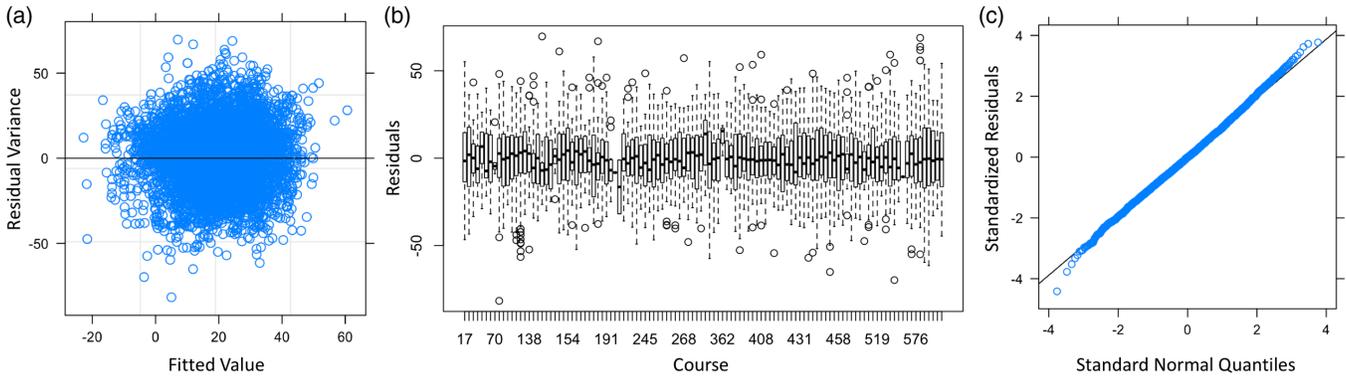

FIG. 3. A visual check of the three assumptions of HLM model 3 from the first MI dataset. (a) Assumption of linearity—A plot of residuals vs fitted values. A random distribution of sample points indicated the model met the assumption of linearity. (b) Assumption of homogeneity of variance—A boxplot of residuals by course. Consistent medians and interquartile ranges indicated the model met the assumption of homogeneity of variance. This assumption was further verified quantitatively using an ANOVA. (c) Assumption of normality—A quantile-quantile plot of the observed and expected values. The points falling close to the line indicated that the model met the assumption of normality.

While the pretest scores are not the focus of this study, they are included because they are strong predictors of student gains and improve the model's fit. We will not, however, substantively discuss them in our interpretation of the models.

We left all other variables in the model uncentered.

#### 4. Assumption checking

We are unaware of any simple method to pool the multilevel models created by multiply imputed datasets in a way that allows for a combined checking of assumptions. For this reason, we performed the assumption checks 10 times, once for each MI dataset. The assumption checks led to similar findings across each version of model 3. For simplicity, we only report the assumption check analysis of the model 3 from the first MI dataset. To test the assumption of linearity, we plotted the residual variance against the fitted values [Fig. 3(a)]. In our visual inspection of Fig. 3(a) we saw no obvious trends and concluded that the model met the assumption of linearity. To test for homogeneity of variance we created a boxplot of the residuals across courses [Fig. 3(b)] and performed an ANOVA of the residuals across courses. A visual inspection of the boxplot showed the courses' residuals had consistent medians and interquartile ranges and therefore met the assumption of homogeneity of variance. The ANOVA supported our visual check because it did not find a statistically significant difference ($p > 0.05$) in the variances across courses. Finally, we visually checked the assumption of normality of residuals using a quantile-quantile plot of the observed and expected values [Fig. 3(c)]. We concluded that the model met the assumption of normality of residuals because of the linearity of the data in the plot. In Sec. III C we pointed out four assumptions for HLM. We did not test the fourth assumption for random intercepts because it was inherently met by our use of HLM.

#### 5. Descriptive statistics

Table V describes the number of students and courses in each category and the mean gains by course (aggregated) and by student (disaggregated). Most of the courses in the data used LAs ($n = 70$) and most of the students in the data were in those LA-using courses ($n = 4100$). The mean gains by course and by student tended to be within a few percentage points with the exception of collaborative courses. The mean gain for all of the students in collaborative courses was 25.0 percentage points. In contrast, the mean gain for the collaborative courses was 15.4 percentage points. We investigated the raw data to understand why this large difference in mean gains occurred.

TABLE V. Counts and mean gains by instrument and course type. Mean gains were examined at the course level through aggregation and student level through disaggregation. Course types were lecture (Lect.), collaborative without LAs (Collab.), and collaborative with LAs (LAs).

|  | Instrument | | | Course type | | |
|---|---|---|---|---|---|---|
|  | Total | FCI | FMCE | Lect. | Collab. | LAs |
| **Course** | | | | | | |
| Count | 112 | 92 | 20 | 18 | 24 | 70 |
| Gain (% pts) | 16.3 | 17.5 | 19.1 | 14.2 | 15.4 | 19.5 |
| St. Dev. | 9.3 | 9.4 | 9.3 | 9.4 | 11.6 | 7.9 |
| **Student** | | | | | | |
| Count | 5959 | 4077 | 1882 | 791 | 1068 | 4100 |
| Gain (% pts) | 18.8 | 19.4 | 19.2 | 12.0 | 25.0 | 19.3 |
| St. Dev. | 21.6 | 21.0 | 22.9 | 19.8 | 21.8 | 21.5 |

Twenty-four courses with 1068 students used collaborative learning. Two of these courses included 248 and 251 students with mean gains of 33.7 and 31.9 percentage points, respectively. The 22 other courses that used collaborative learning included 569 students and the course mean gain for those 22 courses was 12.6 percentage points. These differences illustrate how a few courses with relatively large learning gains and large enrollments can result in large differences between the aggregated and disaggregated means.

The differences between calculating mean gains using aggregation and disaggregation are more stark when examining the standard deviations. The standard deviation of the gains more than doubled in nearly every category when calculated using course (aggregation) versus student gains (disaggregation).

It is not necessarily surprising that the standard deviation of the course data is smaller than the student data. Aggregating the course data reduces the range of values likely to occur, leading to smaller standard deviations for the course level data than the student level data. For example, some students gains were negative, but no course gains were similarly low. Standard deviation (which is the square root of the deviance) plays a central role in many statistical analyses and variations in its value can meaningfully impact any associated findings.

## C. Findings

Table VI shows the coefficients for both the MLR and HLM versions of model 3. Some of the coefficients remained consistent across the models while others varied by large amounts. For example, the coefficients for being in a class with LAs (+7.27 for MLR and +5.94 for HLM) and their statistical significances ($p < 0.001$ for MLR and $p = 0.012$ for HLM) were similar in both models. The coefficient for collaborative learning without LAs, however, differed in both its magnitudes (+12.99 for MLR and

TABLE VI. MLR and HLM coefficients for model 3. Variable labeling is consistent with the HLM model.

| Fixed effect | Final estimation of fixed effects | | | | | |
|---|---|---|---|---|---|---|
| | MLR | | | HLM | | |
| | Coef. | S.E. | $p$ | Coef. | S.E. | $p$ |
| For Intercept 1 $\beta_0$ | | | | | | |
| For Int., 2 $\gamma_{00}$ | 12.01 | 0.95 | <0.001 | 13.76 | 2.14 | <0.001 |
| LA, $\gamma_{01}$ | 7.27 | 1.01 | <0.001 | 5.94 | 2.36 | 0.012 |
| CollNoLA, $\gamma_{02}$ | 12.99 | 1.29 | <0.001 | 3.26 | 3.02 | 0.283 |
| For StudentPre slope, $\beta_1$ | | | | | | |
| For Int., 2 $\gamma_{10}$ | −0.44 | 0.02 | <0.001 | −0.44 | 0.02 | <0.001 |

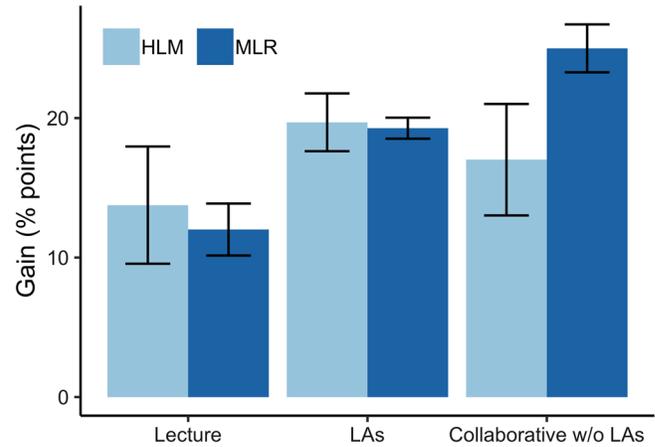

FIG. 4. Predicted gains for average students across course contexts as predicted by HLM and MLR using model 3. Gains were measured as the change in percentage of the questions correct from the pretest to the post-test. Error bars are ±1 standard error. Error bars vary in size based on the number and spread of student scores in each course context.

+3.26 for HLM) and its statistical significances ($p < 0.001$ for MLR and $p = 0.283$ for HLM).

Figure 4 shows the predicted gains for an average student in an average course based on model 3 for both the HLM and MLR analyses. The predicted gains for students in lecture-based and LA-supported courses did not vary much between the MLR and HLM models. The predicted gain for students in collaborative learning courses without LAs, however, went from being 2.1× the gains of lecture-based courses in the MLR model to being 1.2× in the HLM model. The course context with the largest gains also switched from collaborative learning in the MLR model to LA supported courses in the HLM model. As indicated by the error bars in Fig. 4, the difference between predicted gains in collaborative and LA supported courses was statistically significant in the MLR analysis but not in the HLM analysis. These large differences between the MLR and HLM analysis for collaborative courses were largely driven by the two outlier collaborative courses. The two outlier collaborative courses had larger enrollments and larger mean gains than the other 22 collaborative courses. The MLR model ignored the structure of the data and overweighted the importance of the two outlier courses.

## D. Discussion

Our comparison of MLR and HLM with real physics course data demonstrates that the theoretical differences between MLR and HLM can lead to important differences in model findings. In this example, the MLR model predicts that courses that use collaborative learning *without* LAs have the largest student gains. In contrast, the HLM model predicts no statistically significant differences between collaborative with and without LAs. The MLR model also produced artificially small standard errors that were around half the size of those produced by HLM. The difference between the HLM and MLR models resulted from MLR failing to account for two large, outlier courses. These two courses had half of the total students in collaborative courses and had much higher gains than the other 22 collaborative courses. MLR's overconfidence in model precision increases the odds of making type I errors in which coefficients are falsely identified as being statistically significantly different from zero. The conclusions and recommendations that one might make to instructors, administrators, and policy makers greatly differ between the two regression techniques used to analyze this dataset. These analyses demonstrate the importance of using a method designed for modeling hierarchical data to analyze educational data with students nested within courses.

When analyzing multilevel data it is very unlikely that the student data will be independent. Most likely there will be course-level differences that lead student outcomes within a course to be clustered, thereby creating dependencies in the dataset. MLR models do not account for these dependencies because one of the assumptions of MLR is that the data are independent. HLM is designed to address these dependencies by leveraging the group correlations to accurately identify both level-1 effects and level-2 effects. Even if there are no level-2 predictor variables included in a model, HLM can account for the impact of the unknown course differences (e.g., percent majors, time of day, curriculum, and instructor experience). To accomplish this, HLM allows each level-2 group to vary independently. While HLM's flexibility allows it to handle more diverse datasets and create more accurate models it also creates its biggest drawback: lower statistical power.

## V. CONCLUSION

MLR can be a powerful tool for isolating variables of interest while accounting for other variables. MLR models, however, assume that each data point in the dataset is independent from all other data points. This assumption is often not true for datasets with hierarchical data (e.g., students within multiple courses) and can create biased findings. The historical failure to account for the hierarchical structure of the data in many PER studies calls into question the validity and reliability of their claims.

HLM leverages the hierarchical structure of datasets to create more accurate estimates of variable coefficients across levels. HLM has been in use since the mid 1980s [58] and commercial software designed to perform HLM has been available since the 1990s [59]. While the field of PER has not broadly adopted the use of HLM, the use of HLM in the larger field of education research started in the 1990s [60] and has become more commonly used in the intervening decades. HLM models require more data and calculations than MLR models but with the availability of large-scale databases (e.g., DataExplorer [61], E-CLASS [62], and LASSO [36]) and modern computer processors significantly reduce these barriers to use. Physics education researchers will hopefully use the power of these large-scale hierarchical datasets to answer novel research questions. When analyzing these hierarchical datasets, however, it is important that physics education researchers adopt the use of appropriate quantitative methods (i.e., HLM). Analyzing hierarchical datasets with MLR may bias or skew results, and it could take PER researchers extensive effort to undo the harm caused by the findings and implications derived from the use of improper analysis techniques. While no simple, *post hoc* fix exists for studies that used inappropriate modeling techniques, we hope that this article will support future researcher's use of HLM when analyzing hierarchical datasets.

## ACKNOWLEDGMENTS

This work is funded in part by NSF-IUSE Grant No. DUE-1525338 and is Contribution No. LAA-057 of the Learning Assistant Alliance. We are grateful to the Learning Assistant Program at the University of Colorado Boulder for establishing the foundation for LASSO and LASSO studies.